# Grapheayne: a class of low-energy carbon allotropes with diverse optoelectronic and topological properties

Yan Gao[1], Chengyong Zhong[2], Shengyuan A. Yang[3,4], Kai Liu[1]*, and Zhong-Yi Lu[1]*

[1]*Department of physics and Beijing Key Laboratory of Opto-electronic Functional Materials & Micro-nano Devices, Renmin University of China, Beijing 100872, China*
[2]*Institute for Advanced Study, Chengdu University, Chengdu 610106, China*
[3]*Research Laboratory for Quantum Materials, Singapore University of Technology and Design, Singapore 487372, Singapore*
[4]*Center for Quantum Transport and Thermal Energy Science, School of Physics and Technology, Nanjing Normal University, Nanjing 210023, China*

A series of carbon allotropes with novel optoelectronic and rich topological properties is predicted by systematic first-principles calculations. These fascinating carbon allotropes can be derived by inserting acetylenic linkages ($-C \equiv C-$) into graphite, hence they are termed as grapheaynes. Grapheaynes possess two different space groups, $P2/m$ or $C2/m$, and contain simultaneously the *sp*, *sp*$^2$, and *sp*$^3$ chemical bonds. They have formation energies lower than the already experimentally synthesized graphdiyne and other theoretically predicted carbon allotropes with acetylenic linkages. Particularly, when the width *n* of grapheayne-*n* exceeds 15, its cohesive energy is lower than that of diamond, and approaches that of graphite with increasing *n*. Remarkably, we find that some grapheaynes behave as semiconductors with direct narrow band gaps and own the highest absorption coefficients among all known semiconducting carbon allotropes, while some others are topological semimetals with nodal lines. Especially, some grapheaynes can be engineered with tunable direct band gaps in the range of 1.07-1.87 eV and have ideal properties for photovoltaic applications. Our work not only uncovers the unique atomic arrangement and prominent properties of the grapheayne family, but also offers a treasury that provides promising materials for catalyst, energy storage, molecular sieves, solar cell, and electronic devices.

---

Corresponding authors: kliu@ruc.edu.cn; zlu@ruc.edu.cn

Graphene, a two-dimensional (2D) monolayer of graphite with $sp^2$ hybridization, has attracted great interest owing to its unique 2D configuration, outstanding physical and chemical properties[1-3], as well as its promising applications in microelectronics, nanoelectronics, energy, and chemical/biological sensors[4-7]. Nevertheless, broad applications of graphene-based materials are hindered by the vanishing band gap of intact graphene, impacting its direct usage in the integration of semiconductor devices[8, 9]. For this reason, tremendous effort has been devoted to controllable strategies for opening a band gap in graphene[10, 11].

Two interesting approaches have been highlighted. One is to fabricate graphene into the form of nanoribbons[12]. The quantum confinement effect can open a band gap which oscillates with and can be controlled by the width of the ribbon[13, 14]. The other approach is to transform the pure $sp^2$ bonding in graphene by adding other ($sp$ or $sp^3$) bonding characters[15-17]. This can be achieved by chemical functionalization or by utilizing the versatile bonding types of carbon itself[18, 19]. A prominent example is graphyne[20-23], which is a family of 2D planar carbon allotropes composed of benzene rings ($sp^2$ carbons) and acetylenic linkers ($sp$ carbons). The $sp$-$sp^2$ hybrid bonding character can dramatically affect the electronic properties and helps to tune the bandgaps[24, 25]. For instance, the experimentally synthesized graphyne family member, graphdiyne[22], is a semiconductor with a gap ~0.5 eV[16].

In this work, we propose a new strategy for gap engineering in carbon by combining the above two approaches. The idea is to assemble graphene nanoribbons via acetylenic linkages ($-C \equiv C-$) to construct a carbon network structure.

Depending on the width of the nanoribbon, this gives a family of new 3D carbon allotropes, which can be termed as grapheaynes. By first-principles calculations, we show that the grapheaynes enjoy excellent stability. They are energetically more stable than the experimentally synthesized graphdiyne and other proposed carbon allotropes with acetylenic linkages, and they are thermally stable up to 1000 K. Following the design philosophy, the geometric confinement of the nanoribbons and the $sp$-$sp^2$-$sp^3$ hybrid bonding endow the grapheayne family with a variety of electronic band features. We find that if the width of the nanoribbon $n$ satisfies $n = 3p + 2$ ($p$ is an integer), then the system is a topological nodal loop semimetal; otherwise, it is a semiconductor. Interestingly, several semiconducting members of the family (with $n =$ 3, 7, and 10) possess direct band gaps in the range of 1.0-1.5 eV, which are comparable to that of silicon (1.1 eV) and are very close to the optimal band gap of 1.34 eV for solar cell absorber materials. This suggests a great potential of grapheaynes for optoelectronic applications, and our calculation confirms that they have strong optical absorption response superior than other carbon allotropes as well as most optoelectronic materials. In addition, as an advantage of 3D carbon network structures, the grapheaynes are also promising for energy storage and molecular sieve applications. We also discuss possible experimental routes for synthesizing and characterizing these new carbon allotrope materials.

## Results

**Lattice structure and stability.** As we have mentioned, the grapheaynes are a family of 3D carbon allotropes. They can be labeled by the width *n* of the composing graphene nanoribbons in their structures. Two representative examples, grapheayne-4 and 5, are shown in Figs. 1a and 1b, respectively. Their corresponding primitive cells are plotted in Figs. 1c and 1e and Brillouin zones (BZs) given in Figs. 1d and 1f. In the structures, one can recognize three kinds of carbon atoms according to their bonding characters. (1) The carbon atoms connecting the $-\text{C}\equiv\text{C}-$ linkages to the nanoribbons are $sp^3$ hybridized (denoted as C3, cyan color). (2) The atoms in the nanoribbons are $sp^2$ hybridized (denoted as C2, blue color). (3) The atoms on the $-\text{C}\equiv\text{C}-$ linkages are *sp* hybridized (denoted as C1, red color). The ratio between the numbers of *sp* (C1), $sp^2$ (C2), and $sp^3$ (C3) hybridized carbon atoms in grapheayne-*n* is 1: *n*: 1. Thus, the grapheaynes indeed possess an $sp$-$sp^2$-$sp^3$ hybrid carbon bonding character.

The key structural parameters and basic properties of grapheayne-*n* (*n*=1-10) are listed in Table I. For comparison, several other typical 3D carbon allotropes, including carboneyane[26], T-carbon[27], diamond, and graphite, are also listed. For consistency, all data shown here are obtained by our own calculations using the same computational method (see the Methods section).

From Table I, one can see that the grapheayne family has two distinct space groups depending on the parity of *n*: if *n* is odd, it belongs to $P2/m$; if *n* is even, it belongs to $C2/m$. In the fully optimized structures of grapheayne-1 to 10, the coexistence of *sp*, $sp^2$, and $sp^3$ hybridizations lead to different $\text{C}-\text{C}$ bonds with different bond lengths: the C1-C1 bonds (1.21-1.22 Å), the C1-C3 bonds (1.46-1.48

Å), the C2-C2 bonds (1.34-1.46 Å), the C2-C3 bonds (1.52-1.55 Å), and the C3-C3 bonds (1.55-1.63 Å). Due to the insertion of the $-\text{C}\equiv\text{C}-$ linkages, the interlayer spacing between graphene-like sheets increases from 3.35 Å for graphite to 4.12-5.00 Å for the grapheaynes. Meanwhile, the mass density of grapheayne-$n$ decreases from 2.26 g/cm$^3$ to 1.70 g/cm$^3$ with the increasing width $n$, most of which are much smaller than that of graphite. Regarding the bulk modulus, all the grapheaynes have smaller values (ca. 200.51-255.92 GPa) compared with the graphite (266.66 GPa), but still larger than those of carboneyane (155.03 GPa) and T-carbon (155.52 GPa).

Stability is an important issue when proposing a new carbon allotrope. By analyzing the calculated cohesive energies ($E_{\text{coh}}$), one finds that grapheaynes have very low formation energies compared with other systems containing the $-\text{C}\equiv\text{C}-$ linkages. For example, it is more stable than graphyne (-8.58 eV per C atom) which has been predicted to be the most stable among the 2D graphyne family, and is also more stable than carboneyane (-8.27 eV per C atom) which is another metastable 3D carbon allotrope with $sp$-$sp^2$-$sp^3$ chemical bonds. We also calculated the cohesive energies ($E_{\text{coh}}$) as a function of width $n$ for grapheayne-$n$. The result is plotted in Fig. 2. In the figure, the black dotted line represents the energy of diamond and the lower bound of the energy axis represents the cohesive energy of graphite. One can observe that when the width of grapheayne-$n$ exceeds 15, the system is energetically lower than diamond, and its $E_{\text{coh}}$ approaches graphite (Fig. 2). We have also calculated the total energy against volume curves for fourteen typical carbon allotropes, including grapheayne-$n$ ($n$=1-10), carboneyane, T-carbon, diamond, and graphite (as shown in Supplementary Fig. S1). Among them, graphite has the lowest energy, followed by diamond, while most grapheaynes approach diamond. This indicates that the grapheaynes are energetically metastable. It is worth noting that with increasing negative pressure, the grapheaynes become more and more energetically favorable than T-carbon, diamond, and even graphite (see Fig. S1).

To inspect whether grapheaynes are dynamically stable or not, we calculated their phonon spectra. The results for grapheayne-4 and 5 are shown in Figs. 3a and 3b, respectively (others are shown in Supplementary Figs. S2 and S3). One can see that there is no imaginary phonon mode in the whole BZ, which confirms that grapheaynes are dynamically stable. We have also checked the thermal stability of the grapheaynes by performing the *ab initio* molecular dynamics (AIMD) simulations at finite temperatures. For grapheayne-4 (grapheayne-5), we find that they can retain their integrity at 1000 K (1200 K) (Fig. 3c-3f), indicating its excellent stability at high temperature.

**Electronic, optical, and topological properties.** The grapheayne family exhibits rich electronic properties depending on the width of the nanoribbon component. We find that grapheayne-$n$ is a topological nodal-line semimetal (indicated by the red box in Fig. 2) if $n = 3p + 2$ (where p is an integer); otherwise, it is a semiconductor. Among the semiconducting grapheaynes, there are a series of direct narrow band gap semiconductors (the blue box in Fig. 2) as well as indirect-gap semiconductors (the green box in Fig. 2). Let's first consider grapheayne-4, which is an example of a direct gap semiconductor. Fig. 4a presents the band structure of grapheayne-4, which possesses a direct band gap of 1.31 eV (PBE) between the valence band maximum (VBM) and the conduction band minimum (CBM) at the M point of the BZ. This gap value is very close to the optimal band gap of 1.34 eV for solar cells. We have also performed calculations with the hybrid functional approach (HSE06), which yields a slightly larger band gap of 1.87 eV (Supplementary Fig. S7a). The band structure of grapheayne-4 exhibits large dispersions around CBM and VBM, therefore, relatively low effective masses for electrons and holes and large intrinsic mobilities are expected. Figure 4c shows that the grapheayne-3, 4, 7, and 10 have effective masses of electrons and holes comparable to some well-known electronic materials, such as GaN, ZnO, Si, and carbon kagome lattice (CKL)[28]. Especially, due to the graphene-like nanoribbon unit in grapheaynes, the effective masses of grapheayne-4 and grapheayne-7 in the ribbon plane (i.e., $m_e^\parallel$ and $m_{hh}^\parallel$) are very small, which may lead to promising potential application in electronic devices.

Besides grapheayne-4, there are also a series of grapheaynes with direct and narrow band gaps (Table I), such as grapheayne-3, grapheayne-7, and grapheayne-10, whereas the CBM and VBM are located at B, A, and M, respectively (Supplementary Figs. S4c, S5a, and S5d). The PBE gaps of them are 0.62, 0.82, and 0.71 eV, respectively, while the corresponding hybrid functional gaps are 1.12, 1.19, and 1.07 eV (Supplementary Figs. 8a, 7d, and 8b). According to the Shockley-Queisser limit[29], for suitable solar cell materials, the band gap value should lie in the range of 1.0-1.5 eV, in order to maximize energy conversion efficiency. Therefore, our result implies that grapheaynes with tunable direct band gaps could be a good candidate for photovoltaic applications.

To gain better understanding of the band structure, in Fig. 4b, we plot the partial density of states (PDOS) of grapheayne-4. We note that the *p* orbital of C2 atoms dominate the bands near the band gap (the red-colored bands in Fig. 4a). In contrast, the *p* orbitals of C1 atoms occupy the lower valence bands (the blue-colored bands in Fig. 4a), while the contribution from the C3 atoms are insignificant at low energy. In order to further investigate the orbital characteristics at the band edge, we calculate the charge densities for VBM and CBM states. The result shown in Fig. 4a indicates that the VBM is composed of $\pi$ states while the CBM is made up of $\pi^*$ states, quite similar to graphene. It is noteworthy that the optical dipole transition between the $\pi$ and $\pi^*$ states is allowed, thus the material should have a strong optical absorption capability. Hence we calculate the imaginary part of dielectric function $\varepsilon_2(E)$ (with the hybrid functional approach) to estimate the optical properties of grapheayne-3, 4, 7, and 10. It can be seen from Fig. 4d that the optical absorptions start at the direct gap transition energies. Remarkably, the calculated optical absorptions of these grapheaynes are much stronger than those of GaN, ZnO, CKL, and T-carbon. Particularly, to the best of our knowledge, grapheayne-10 possesses the highest absorption coefficients among all semiconducting carbon allotropes. This indicates that grapheayne-3, 4, 7, and 10 can be well used in photovoltaic devices. In addition, the absorption of grapheaynes starts at different energies (depending on *n*) that cover a range of frequencies in the solar spectrum. Thus, an assembly of these structures would absorb even more sunlight.

On the other hand, in the grapheayne family, there are a series of topological nodal-line semimetals. We take grapheayne-5 as an example to illustrate the key physics. Since the effect of spin-orbital coupling (SOC) for carbon is negligible, we calculated the orbital-projected band structure of grapheayne-5 (shown in Fig. 5a) without SOC. As can be seen, the valence and conduction bands of grapheayne-5 cross at about -73 meV below the Fermi level, exhibiting linear dispersion along the high-symmetry *B-A* path (along the direction of zigzag-edged graphene nanoribbons) in the bulk BZ (see the inset in Fig. 5a). To further explore the electronic characteristics of grapheayne-5, we calculated its partial density of states (PDOS). As shown in Fig. 5b, the $p_x$ orbital of C2 atoms dominate the energy bands near the Fermi level, corresponding to the red bands in Fig. 5a, while the contributions of other orbitals of C1, C2, and C3 atoms are insignificant. One can see that the density of states is nearly zero around the Fermi level, exhibiting a semimetallic behavior. We also examined the band structure by using the hybrid functional as shown in Supplementary Fig. S7b. Analyzing the charge density of two states ($B_1$ and $B_2$) around the crossing point reveals that the $p_x$ orbital of C2 atoms form π bands (see the inset in Fig. 5a), which is very similar to the origin of Dirac cone of graphene, whose π bands are attributed to the $p_z$ orbitals. Therefore, we can use a simple tight-binding (TB) model only consisting of C2 atoms to capture the essential physics around the Fermi level for grapheayne-5 (refer to the Supplementary Information with Fig. S9).

According to the crystal symmetry, grapheayne-5 has the little point group of $C_2$

along the *B-A* path, and the two bands involved in the crossing belong to different irreducible representations of A and B along the path (see the inset in Fig. 5a). As a result, the band crossing is protected by symmetry. In addition, the system preserves the inversion (*P*) and time reversal (*T*) symmetries. For such *PT*-symmetric system without SOC, the crossing point cannot be an isolated single point. Indeed, a careful inspection reveals that the band crossing points form a continuous nodal ring in the whole BZ (see the inset in Fig. 5a). This nodal ring is protected by a quantized one-dimensional winding number $N = \frac{1}{\pi}\oint_l \text{Tr}[A_k] \cdot dk$ where $A_k$ is the Berry connection at point $k$ for the occupied bands, and the integration path $l$ is around a loop encircling the ring. This winding number is essentially the Berry phase in unit of π. Our calculation confirms that $N = 1$ for the nodal ring.

One important characteristic of a topological nodal-ring semimetal is the presence of "drumhead-like" surface states. Here, we study the surface spectrum (Figs. 5c and 5d) of grapheayne-5 on the (001) surface. The result confirms that the "drumhead-like" surface states exist and they are nestled outside the projected nodal ring in Fig. 5d (the red dashed lines). These "drumhead-like" surface states are slightly below the Fermi level; hence they should be easily detected by angle-resolved photoemission spectroscopy (ARPES). For grapheayne-2 and grapheayne-8, they also possess nodal rings, and the patterns of the nodal ring and surface states are different (see Supplementary Fig. S6). In particular, grapheayne-2 has two helical nodal loops in the first BZ. Like grapheayne-5, their nodal lines are also protected by *PT* symmetry with a nontrivial winding number.

## Discussion

In this work, we have proposed a strategy for engineering the band gap in carbon, namely, by assembling graphene-like nanoribbons with acetylenic linkages, which combines the quantum confinement effect and hybrid bonding character to achieve a strong band structure modulation. According to above strategy, we design a family of novel stable carbon allotropes — the grapheaynes, with fascinating electronic properties.

We have shown that the semiconducting grapheayne members can have direct band gaps in the range of 1.07 to 1.87 eV, which overlaps with the desired range for solar cell materials. Indeed, the Shockley-Queisser limit[29] suggests that the theoretically maximum solar converting efficiency 33.7% of a single-junction solar cell occurs for a semiconductor with a band gap of 1.34 eV, as evidenced by silicon[30,31], GaAs[32], CdTe[33], InP[34], $CuIn_xGa_{1-x}Se_2$[35], and hybrid organic-inorganic perovskite compounds[36] ($CH_3NH_3PbX_3$, X = I, Br and Cl). Nevertheless, these materials proposed so far have various limitations. For example, silicon is an indirect-gap (1.1 eV) semiconductor and its optical absorption is limited by the requirement of phonon assistance. As, Cd, and Pb are toxic; In is a rare element; and $CH_3NH_3PbX_3$ is not stable[37]. In contrast with these well-known materials, the grapheaynes may be better candidates, since they are non-toxic, consist of carbon which is cheap and abundant, and most importantly, their band gaps might be adjusted.

To facilitate experimental observation, we simulate the the X-ray diffraction (XRD) patterns of the grapheaynes along with several other carbon allotropes. The results are given in Supplementary Fig. S10, which can be used as structural indicators in future experiments. Taking grapheayne-4 as an example, we also suggest a possible experimental scheme for synthesis (Supplementary Fig. S11). One may start from the graphene, and functionalize the graphene with bromide. Then, stacking the multilayer brominated graphene sheets and inserting acetylene molecules and finally deacidification will lead to the formation of grapheayne-4. Given that grapheaynes are both energetically and dynamically stable and in consideration of the rapid development and maturity in experimental technology, we expect that grapheaynes can be synthesized in the near future.

Such a series of grapheaynes, once obtained, would have several other advantages and promising applications. First, the interlayer spacing increases from 3.35 Å in graphite to 4.12-5.00 Å in grapheaynes due to the insertion of acetylenic linkages ($-C \equiv C-$), which may provide reversible storage space and storage capacity for lithium and sodium ions. In fact, these acetylenic linkages also enhance the stability of grapheaynes, which may solve the swelling problem as in lithium ion batteries. Second, grapheaynes have the graphene-like basic unit, so they may inherit the excellent properties of graphene, such as super-high conductivity and super-high thermal conductivity. In particular, the grapheayne family also contains a series of semiconductors with tunable direct band gaps in the range of 1.07 to 1.87 eV, which may make the graphene-based materials be used in semiconductor electronic devices and photovoltaic.

## Methods

**Computational methods:** The first-principles electronic structure calculations were based on density functional theory (DFT) with the Perdew-Burke-Ernzerhof (PBE) approximation[38] to the exchange-correlation energy. The core-valence interactions were described by the projector augmented wave (PAW) method[39], as implemented in the VASP package[40]. The kinetic energy cutoff of 550 eV was adopted for the plane wave basis. The atomic positions were fully relaxed by the conjugate gradient method; the energy and force convergence criteria were set to be $10^{-6}$ eV and $10^{-3}$ eV/Å, respectively. The *k*-point meshes $7 \times 7 \times 7$ and $9 \times 7 \times 11$ were used for the Brillion zone (BZ) integration of grapheayne-4 and grapheayne-5, respectively. The band gaps of semiconductors were also calculated by using the Heyd-Scuseria-Ernzerhof (HSE06) hybrid functional[41]. The frequency-dependent dielectric matrix was calculated using the method described by Gajdos et al. within PAW potentials[42]. In order to study the dynamical stability, we used a finite displacement approach as in the PHONOPY package[43]. The thermal stability was investigated with the *ab initio* molecular dynamics (AIMD) simulations in a canonical ensemble[44] with a Nose-Hoover thermostat. The band crossing characteristics were analyzed with the WannierTools package[45] according to the tight-binding model constructed via the Wannier90 code[46].

## Acknowledgements:

We wish to thank Peng-Jie Guo and Weikang Wu for helpful discussions. This work was supported by the National Key R&D Program of China (Grant No. 2017YFA0302903), the National Natural Science Foundation of China (Grants No. 11774422, 11774424 and 11804039), and the Singapore Ministry of Education AcRF Tier 2 (MOE2017-T2-2-108). Computational resources were provided by the Physical Laboratory of High Performance Computing at Renmin University of China.

**Author contributions:**

Y.G. proposed the grapheaynes. Y.G., K.L. and Z.-Y.L. conceived the original ideas. Y.G. and C. Z. conducted the calculations. Y.G., K.L., Z.-Y.L., and S.A.Y. wrote the manuscript. All authors discussed the results and commented on the manuscript at all stages.

**Additional information:**

**Competing financial interests:** The authors declare no competing financial interests.

# Figures and Tables

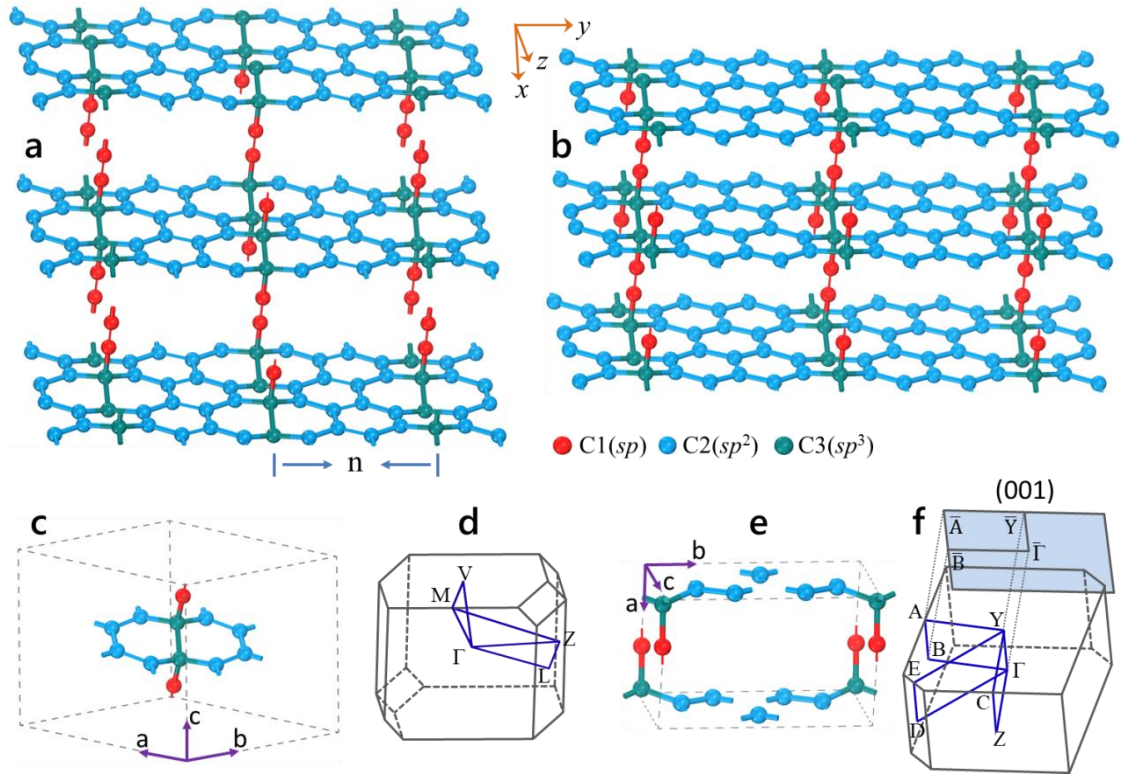

**Figure 1. Crystal structure and Brillion zone (BZ) of grapheayne-n.** The optimized atomic structure of (a) grapheayne-4 and (b) grapheayne-5, both of which share the basic structural units of graphene and acetylenic linkages $-C\equiv C-$. They belong to the same series of grapheayne-n, formed by $-C\equiv C-$ linking the armchair nanoribbons of width n. The red C1, blue C2, and cyan C3 balls are $sp$, $sp^2$, and $sp^3$ atoms, respectively. (c) Primitive cell and (d) the BZ of grapheayne-4. (e) Primitive cell and (f) the BZ of grapheayne-5 with the projected two-dimensional BZ of the (001) surface.

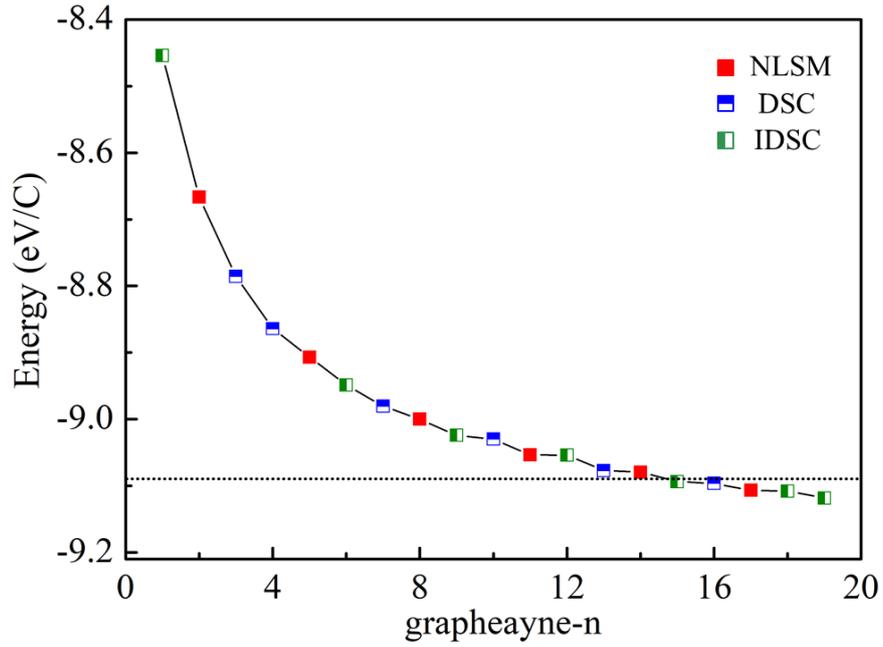

**Figure 2. The cohesive energies ($E_{coh}$) as a function of width n for grapheayne-n.** The black dotted line represents the energy of diamond and the lower bound of the energy axis represents the cohesive energy of graphite. When the width of n exceeds 15, the $E_{coh}$ of grapheaynes will be lower than that of diamond, and then approach graphite. The red, blue and green boxes represent topological nodal-line semimetal (NLSM), direct band gap semiconductor (DSC) and indirect band gap semiconductor (IDSC), respectively.

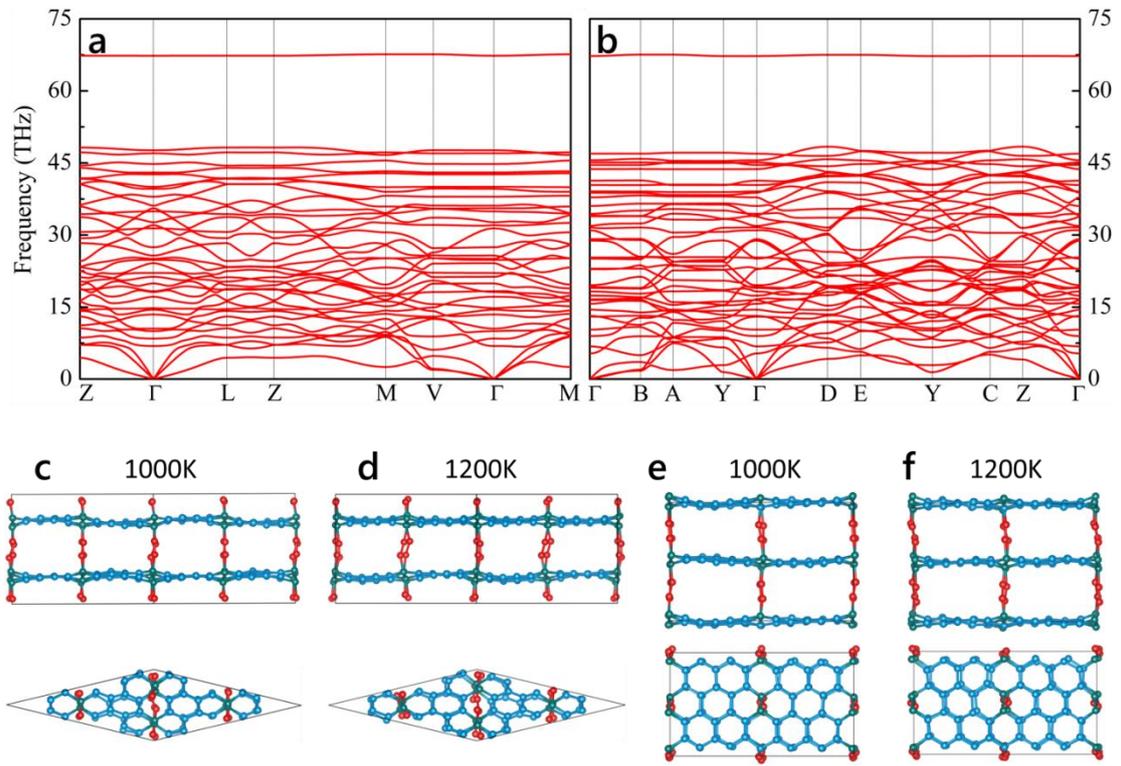

**Figure 3. Dynamical and thermal stability.** Phonon dispersions of (a) grapheayne-4 and (b) grapheayne-5 in the whole BZ. (c-f) Respective top and side views of the snapshots for the equilibrium structures of grapheayne-4 and grapheayne-5 at the temperatures of 1000 K and 1200 K after 15-ps *ab initio* molecular dynamics simulations.

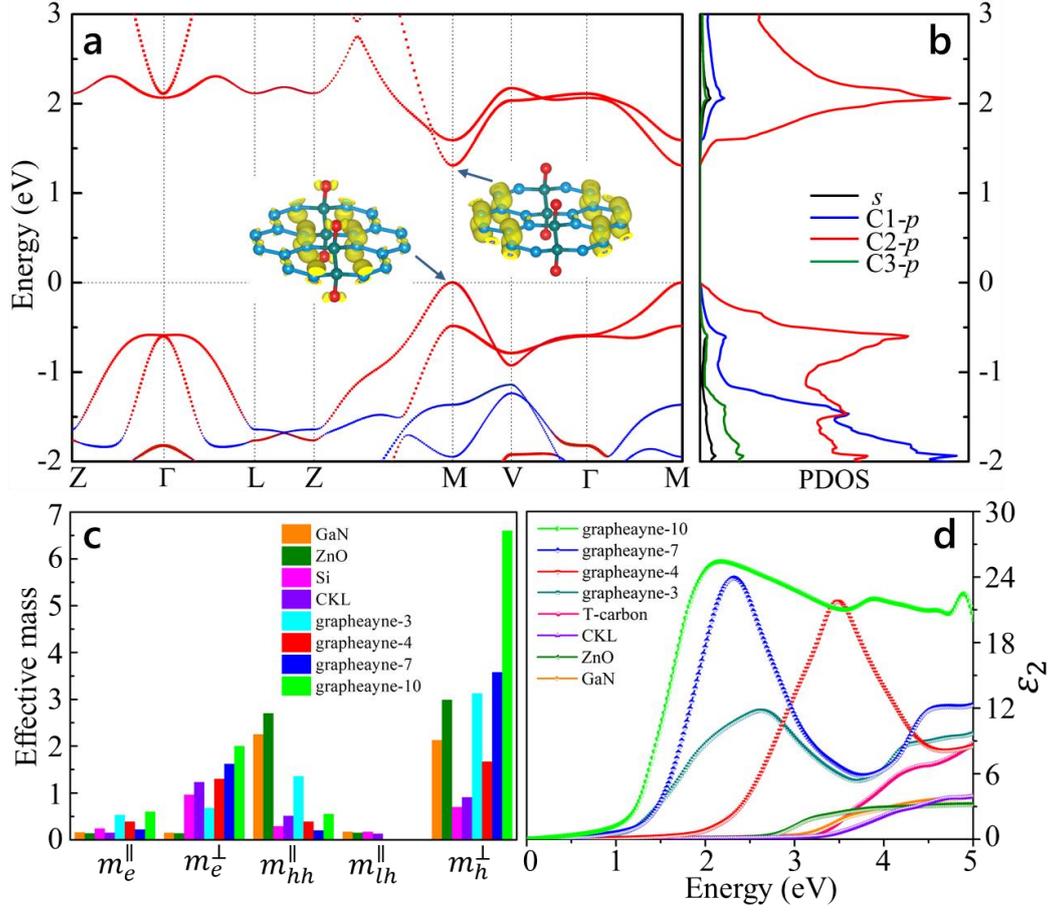

**Figure 4. Electronic and optical properties.** (a) Band structure of grapheayne-4. Insets show the side views of charge densities at valence band maximum and conduction band mimimum. (b) Partial density of states (PDOS) of grapheayne-4, in which the blue, red, and green lines represent the $p$ orbitals of C1, C2, and C3 atoms, respectively. (c) Effective mass of GaN, ZnO, Si, CKL, grapheayne-3, 4, 7, and 10 calculated with the Perdew-Burke-Ernzerhof (PBE) functional. For grapheayne-3, 4, 7, and 10, $m_e^\parallel$ ($m_{hh}^\parallel$) and $m_e^\perp$ ($m_h^\perp$) are the effective masses of electrons in graphene-like plane (light holes) and normal to graphene-like plane (heavy hole), respectively. For hexagonal GaN, ZnO, and CKL, $m_e^\parallel$ ($m_{hh}^\parallel$ and $m_{lh}^\parallel$) and $m_e^\perp$ ($m_h^\perp$) are the in-plane effective masses of electrons (heavy and light holes) and c-axis effective masses of electrons (heavy hole), respectively. For Si, $m_e^\parallel$ and $m_e^\perp$ are the longitudinal and transverse masses of electrons, $m_{hh}^\parallel$ and $m_{lh}^\parallel$ are the effective masses of heavy and light holes along [001], and $m_h^\perp$ is the effective mass of heavy hole along [111]. (d) Imaginary part of dielectric functions $\varepsilon_2$ as a function of energy for GaN, ZnO, CKL, T-carbon, grapheayne-3, 4, 7, and 10 calculated with the hybrid functional.

**22 / 24**

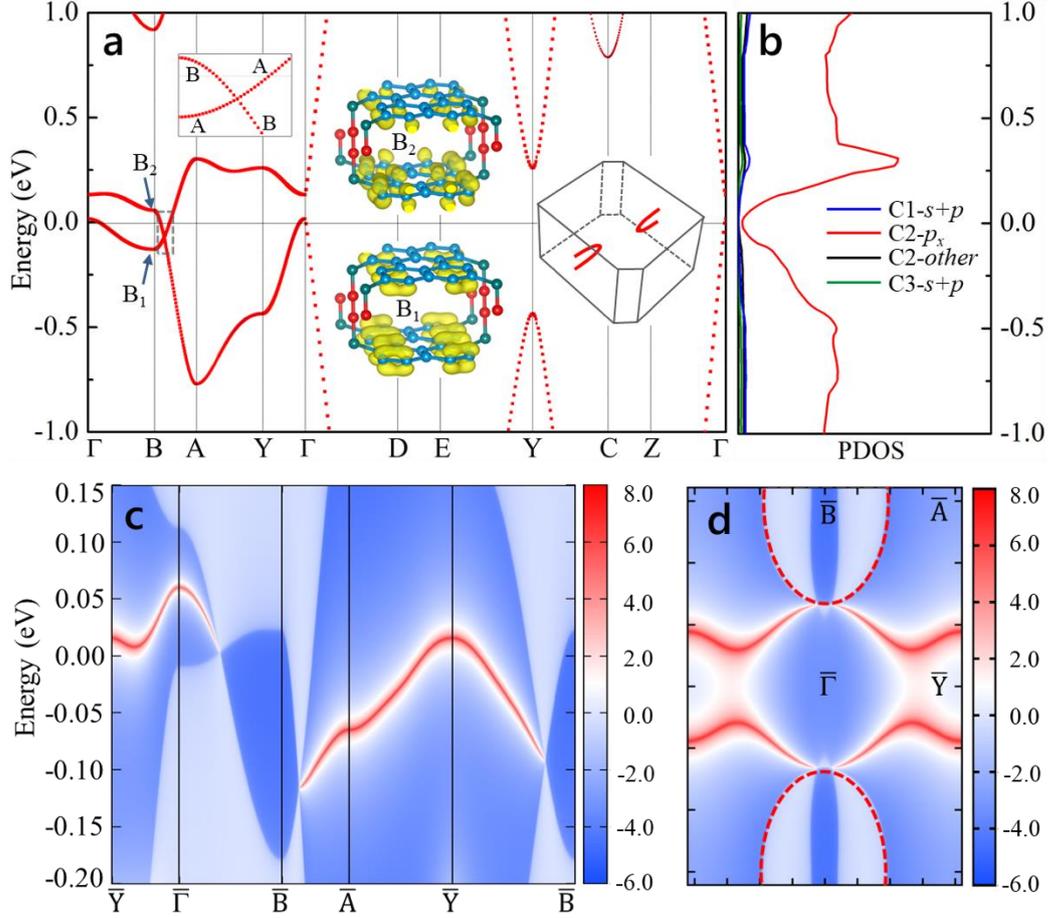

**Figure 5. Band structure of grapheayne-5 and its topological state.** (a) Projected band structure of grapheayne-5, where the red bands correspond to the $p_x$ orbital of C2 atoms. The orbits of other carbon atoms are insignificant around the Fermi level. Inset shows the side views of charge densities at $B_1$ and $B_2$ points and the perspective view of the topological nodal ring in the first BZ, respectively. A and B indicate the irreducible representations of the band crossing point along the *B-A* path. (b) The PDOS for grapheayne-5, in which the states around the Fermi level are contributed by the $p_x$ orbital of C2 atoms. (c) Surface energy bands and (d) surface spectrum at a fixed energy $E - E_f = 0$, where the red dashed lines show the projection of the red nodal ring on the (001) surface in Fig. 5a.

| Structures | Space groups | Lattice parameters(Å) | | | Angles (°) | | | Densities (g/cm³) | Bond lengths(Å) | Bulk moduluses (GPa) | $E_{coh}$ (eV/C) | Ratios | $E_{gap}$ (eV) | Types |
|---|---|---|---|---|---|---|---|---|---|---|---|---|---|---|
| | | a | b | c | α | β | γ | | | | | | | |
| grapheayne-1 | $P2/m$ | 4.91 | 2.62 | 4.12 | 90.00 | 90.76 | 90.00 | 2.26 | 1.21-1.63 | 255.92 | -8.45 | 1:1:1 | 0.16 | IDSC |
| grapheayne-2 | $C2/m$ | 4.31 | 4.31 | 6.04 | 107.23 | 107.23 | 120.34 | 2.05 | 1.22-1.57 | 234.32 | -8.67 | 1:2:1 | ~ | NLSM |
| grapheayne-3 | $P2/m$ | 4.90 | 5.00 | 4.29 | 90.00 | 99.58 | 90.00 | 1.93 | 1.22-1.55 | 222.91 | -8.79 | 1:3:1 | 0.62 | DSC |
| grapheayne-4 | $C2/m$ | 6.59 | 6.59 | 6.06 | 101.21 | 101.21 | 142.17 | 1.85 | 1.22-1.56 | 215.88 | -8.86 | 1:4:1 | 1.31 | DSC |
| grapheayne-5 | $P2/m$ | 4.89 | 7.47 | 4.28 | 90.00 | 98.09 | 90.00 | 1.81 | 1.22-1.56 | 212.22 | -8.91 | 1:5:1 | ~ | NLSM |
| grapheayne-6 | $C2/m$ | 8.95 | 8.95 | 6.01 | 98.14 | 98.14 | 152.30 | 1.77 | 1.22-1.56 | 207.77 | -8.95 | 1:6:1 | 0.42 | IDSC |
| grapheayne-7 | $P2/m$ | 4.89 | 9.93 | 4.28 | 90.00 | 98.09 | 90.00 | 1.74 | 1.22-1.56 | 205.59 | -8.98 | 1:7:1 | 0.82 | DSC |
| grapheayne-8 | $C2/m$ | 11.38 | 11.38 | 6.04 | 96.44 | 96.44 | 158.36 | 1.73 | 1.22-1.56 | 203.48 | -9.00 | 1:8:1 | ~ | NLSM |
| grapheayne-9 | $P2/m$ | 4.89 | 12.39 | 4.28 | 90.00 | 98.29 | 90.00 | 1.71 | 1.22-1.56 | 202.03 | -9.02 | 1:9:1 | 0.26 | IDSC |
| grapheayne-10 | $C2/m$ | 13.80 | 13.80 | 6.04 | 95.32 | 95.32 | 162.16 | 1.70 | 1.22-1.56 | 200.51 | -9.03 | 1:10:1 | 0.71 | DSC |
| carboneyane | $C2/m$ | 5.08 | 5.08 | 4.96 | 91.54 | 91.54 | 60.38 | 1.43 | 1.22-1.54 | 155.03 | -8.27 | 1:2:1 | ~ | NLSM |
| T-carbon | $Fd\bar{3}m$ | 5.31 | 5.31 | 5.31 | 60.00 | 60.00 | 60.00 | 1.50 | 1.42-1.50 | 155.52 | -7.92 | ~ | 2.23 | DSC |
| diamond | $Fd\bar{3}m$ | 2.53 | 2.53 | 2.53 | 60.00 | 60.00 | 60.00 | 3.50 | 1.55 | 422.52 | -9.09 | ~ | 4.12 | IDSC |
| graphite | $P6_3/mmc$ | 2.46 | 2.46 | 6.80 | 90.00 | 90.00 | 120.00 | 2.24 | 1.42 | 266.66 | -9.21 | ~ | ~ | M |

**Table I. Comparison between grapheayne-1 to 10 and other representative carbon allotropes.** Space groups, lattice parameters (Å), angles (°), densities (g/cm³), bond lengths (Å), bulk moduluses (GPa), cohesive energies $E_{coh}$ (eV/B), ratios ($sp$:$sp^2$:$sp^3$), band gaps $E_{gap}$ (eV), and material types (indirect band gap semiconductor (IDSC), topological nodal line semimetal (NLSM), direct band gap semiconductor (DSC), and metal (M)) for grapheayne-1 to 10, carboneyane, T-carbon, diamond, and graphite.